\documentclass[hidelinks,prd,eqsecnum,amsmath,amssymb,nofootinbib,12pt,superscriptaddress]{revtex4-2}
\usepackage{dcolumn}
\usepackage{bm}
\usepackage{xcolor}
\usepackage[applemac]{inputenc}
\usepackage[spanish,english]{babel}
\usepackage{amsmath,amssymb,amsfonts,latexsym,cancel,amsthm,hyperref,bbm}
\usepackage{graphicx}
\usepackage{color}
\usepackage{soul}
\usepackage{ulem}
\usepackage{slashed}
\usepackage{braket}
\usepackage[mathscr]{euscript}
\usepackage{physics}
\usepackage{mathrsfs}
\usepackage{pgfplots}
\usepackage{float}
\usepackage{xfrac}
\usepackage{xcolor}
\usepackage{trimclip}
\usepackage{amsmath}
\usepackage{accents}
\definecolor{blue(pigment)}{rgb}{0.2, 0.2, 0.6}
\allowdisplaybreaks

\newcommand{\la}{\langle}
\newcommand{\ra}{\rangle}
\newcommand{\w}{\omega}

\newcommand{\be}{\begin{equation}}
\newcommand{\ee}{\end{equation}}
\newcommand{\bea}{\begin{eqnarray}}
\newcommand{\eea}{\end{eqnarray}}
\newcommand{\bes}{\begin{subequations}}
\newcommand{\ees}{\end{subequations}}

\begin{document}
  \title{\bf Relaxation of Quantum Observables in a 4D Black Hole that forms from Collapse of a Null Shell}
 
    \author{Shohreh Gholizadeh Siahmazgi}\email{gholizs@wfu.edu}
    \affiliation{Department of Mathematics, Wake Forest University, Winston-Salem, North Carolina, 27109, USA}
    \affiliation{Department of Physics, Wake Forest University, Winston-Salem, North Carolina, 27109, USA}
    
\author{An\i l Zengino\u{g}lu}\email{anil@umd.edu} 
\affiliation{Institute for Physical Science and Technology, University of Maryland, College Park, Maryland 20742, USA}

   \author{Paul R. Anderson}\email{anderson@wfu.edu}
    \affiliation{Department of Physics, Wake Forest University, Winston-Salem, North Carolina, 27109, USA}

\author{Alessandro Fabbri}\email{afabbri@ific.uv.es}
    \affiliation{Departamento de F$\acute{\imath}$sica Te\'orica and IFIC, Universidad de Valencia-CSIC,\\
Avenida Vicent Andr\'es Estell\'es 19, 46100 Burjassot, Spain}

\begin{abstract}

   The vacuum polarization and luminosity are computed in a four-dimensional spacetime where a spherically symmetric null shell contracts to form a black hole. This is done by computing the contribution of the spherically symmetric modes of a massless minimally coupled scalar field to the quantities $\la \phi^2 \ra_{\rm in} - \la \phi^2 \ra_U$ and $L_{\rm in} - L_U$.  Here the subscript ``in'' denotes the {\it in} vacuum state for the null shell spacetime and the subscript ``U''  denotes the Unruh state in Schwarzschild spacetime.  The computation is done outside the null shell and outside the event horizon where the spacetime geometry is the Schwarzschild geometry.  It is found that, as expected, these differences approach zero at late times.  However, they do so significantly more slowly than in the two-dimensional case where there is no scattering of the mode functions. In both cases, the difference is dominated by a quasinormal mode at intermediate times and a tail term at late times that decays quartically in time. For the luminosity difference, an analytic calculation verifies the late-time result for the tail term.
     \end{abstract}

    \date{\today}
    \maketitle

\newpage

\section{Introduction}

One of the impressive aspects of Hawking's original calculation of black hole evaporation~\cite{Hawking-75} was that its main prediction of a flux of radiation in a thermal distribution at future null infinity at late times, is independent of how the black hole forms.  Unruh's discovery~\cite{Unruh-1976} of a state for an eternal black hole which has the same flux of radiation at future null infinity for all times has often been used to study quantum effects in black hole spacetimes because it is much easier mathematically to work with eternal black hole spacetimes than ones that form from collapse.  Both types of calculations are extremely important, but neither gives a complete picture.  To fully understand quantum effects in black hole spacetimes one must also consider effects related to the formation of the black hole.  Interesting questions that can be addressed include how quickly the Unruh state is approached and how sensitive are quantities such as the vacuum polarization and the stress-energy tensor to the details of the collapse process.  

In this paper, we investigate the behaviors of the vacuum polarization $\la \phi^2 \ra$, and the luminosity $L$, for a massless minimally coupled scalar field in a spacetime in which a spherically symmetric massless shell contracts to form a black hole. The geometry inside the shell is that of flat space and outside the shell is Schwarzschild.  To our knowledge no previous numerical calculations of these quantities in the case of collapse to a black hole in four dimensions, have been done which used exact numerical solutions to the mode equation.  Prior full numerical calculations of both the vacuum polarization~\cite{Fawcett-Whiting,Candelas-Howard,Candelas-Jensen,And-phi2,Levi-Ori-phi2-1,Levi-Ori-phi2-2,Lanir-Levi-Ori-phi2-inside} and the stress-energy tensor for a scalar field in Schwarzschild spacetime have been done~\cite{Fawcett,Howard-Candelas,Howard,Jensen-McLaughlin-Ottewill-1,Jensen-McLaughlin-Ottewill-2,AHS-1,AHS-2,And-Balbinot-Fabbri,Levi-Ori-Tab,Taylor-Breen-Ottewill}.  Also, the vacuum polarization and the energy flux have been computed close to the shell trajectory for the case of a 4D black hole that is formed from the collapse of a spherically symmetric null shell~\cite{Ori-Zilberman}. 

In~\cite{quasinormal} it was shown for a massless minimally coupled scalar field that some of the mode functions for one formulation of the Unruh state and the difference between the mode functions for the {\it in} vacuum state and the rest of the mode functions for the Unruh state are dominated by a quasinormal mode at intermediate times and a tail term at late times for a fixed radial position.  
Here, we find the same qualitative behavior for both the vacuum polarization at finite radial locations and the luminosity.  However, the tail term is found
to decay as the fourth power in time everywhere. In contrast, for classical black hole perturbations the tail term has previously~\cite{gcpp} been found to decay
cubically at finite radii and quadratically at future null infinity. 

Specifically, we calculate the contribution from spherically symmetric modes to the vacuum polarization and luminosity.  These are often the modes that make the dominant contribution for a spherically symmetric spacetime.  We use two completely different methods.  One is given in~\cite{methods-paper} and involves working in the frequency domain. The modes for the {\it in} state and some of the modes for the Unruh state are expanded in terms of complete sets of mode functions that can be computed using separation of variables.   The other involves working in the time domain.  In this case the modes are first expanded in terms of spherical harmonics, and then the resulting partial differential equation is solved numerically. It is shown that both methods give the same results.  

The unrenormalized expressions for the vacuum polarization is ultraviolet divergent.  However, this divergence is local.  Thus in the region outside the shell the difference between the unrenormalized expressions for the vacuum polarization for the {\it in} and Unruh states, $\Delta \la \phi^2 \ra = \la \phi^2 \ra_{\rm in} - \la \phi^2 \ra_U $, is finite.  Since we are primarily interested in the approach to the Unruh state this is also the quantity of most interest.  The renormalized vacuum polarization for the Unruh state has been computed in~\cite{Levi-Ori-phi2-1,Levi-Ori-phi2-2}.  Thus one could add that expression to our result to find an approximate contribution of the spherically symmetric modes to the vacuum polarization for the {\it in} state.  Similarly, we compute $\Delta L=L_{\rm in}-L_U$, where $L$ denotes the outward flux of Killing energy at future null infinity.

\section{Background}

We work in a 4D spherically symmetric spacetime where a null shell collapses to form a black hole.  
Inside the shell the space is flat with the metric
\be ds^2 = -dt^2 + dr^2 + r^2 d\Omega^2 \;. \ee
The radial ingoing and outgoing null coordinates are
\be u = t - r \;, \qquad  v = t + r \;. \ee
Outside the shell the geometry is the Schwarzschild geometry with the metric
\be ds^2 =  \left( 1 - \frac{2 M}{r} \right) (-dt_s^2 + dr_*^2) + r^2 d\Omega^2 \;, \ee
with $M$ the mass of the black hole and 
\be r_* = r + 2 M \log\left(\frac{r-2M}{2M} \right) \;. \ee
The radial ingoing and outgoing null coordinates are
\be u_s = t_s - r_* \;, \qquad v = t_s + r_* \;. \ee

The coordinates $r$ and $v$ are continuous across the shell.  The shell collapses along the trajectory $v = v_0$.  One finds~\cite{Masser-Parentani,Sandro-Book,Good-Anderson-Evans} that at the surface of the shell
\bea 
    u_s &=& u - 4 M \log \left( \frac{v_H - u}{4M}\right) \;,  \nonumber \\
    u &=& v_H - 4 M W \exp \left(\frac{v_H-u_s}{4M} \right) , \label{u-us-def}
\eea
where $v_H =v_0-4M$ and $W$ is the Lambert W function.

\subsection{Solutions to the mode equation for the {\it in} and Unruh states}
\label{sec:chi-solns}

We focus on the region outside the null shell and outside the future horizon, where as discussed in~\cite{methods-paper}, the modes for the quantum field can be expanded in terms of complete sets of solutions to the mode equation in Schwarzschild spacetime.
The mode equation for a massless minimally coupled scalar field is $\Box f = 0$.  The solutions that are used for the calculations in this paper are all of the form
\be f_{\w \ell m}(t_s, r, \theta,\phi) = \frac{Y_{\ell m}(\theta, \phi)}{r \sqrt{4 \pi \w}} \psi_{\w \ell}(t_s,r) \;, \ee
with 
\bes \bea &&  \left[ -\frac{\partial^2 }{\partial t_s^2} + \frac{\partial^2 }{\partial r_*^2} - V_{\rm eff}(r) \right] \psi_{\w \ell} = 0 \;, \label{psi-eq} \\
&& V_{\rm eff} = \left(1 - \frac{2M}{r} \right) \left[ \frac{2M}{r^3} + \frac{\ell(\ell+1)}{r^2} \right]  \;. \label{Veff-def} \eea \ees

The state for the field in the collapsing null shell spacetime is the natural vacuum state which we call the {\it in} state.
The complete sets of solutions used for the expansions of the modes in the {\it in} state and the Unruh state are built from solutions for which 
\be \psi_{\w \ell} = e^{-i \w t_s} \chi_{\w \ell}(r) \;. \label{psi-Boulware} \ee
These solutions pose a two-channel scattering problem with the incoming basis at past null infinity and the past horizon, and the outgoing basis at future null
infinity and the future horizon.
 The radiative data at these null boundaries have the form
\bea \psi^{\mathscr{I}^-} &=& e^{-i \w v}  \;, \nonumber \\
    \psi^{\mathscr{I}^+} &=& e^{-i \w u_s}  \;, \nonumber \\
    \psi^{\mathscr{H}^-} &=& e^{-i \w u_s}  \;, \nonumber \\
    \psi^{\mathscr{H}^+} &=& e^{-i \w v} \;, \label{Boulware-initial-data}  \eea

Numerical computations of these modes involve solving the equation
\be \left[ \frac{d^2 }{d r_*^2} + \w^2 - V_{\rm eff}(r) \right] \chi_{\w \ell} = 0 \;. \label{chi-eq} \ee
It is useful to consider two solutions to this equation, which in the limit $r \to \infty$ have the following behavior
\bea \chi^\infty_R &\to & e^{i \w r_*}  \;, \nonumber  \\
    \chi^\infty_R &\to & e^{-i \w r_*} \;. \label{chi-inf-def} \eea
Since $V_{\rm eff} = 0$ at the horizon, these take the general form there
\bea \chi^\infty_R & \to & E_R e^{i \w r_*} + F_R e^{-i \w r_*} \;, \nonumber \\
     \chi^\infty_L & \to & E_L e^{i \w r_*} + F_L e^{-i \w r_*} \;. \label{near-horizon} \eea
The coefficients $E_R, \; F_R, \; E_L$, and $\; F_L$ are computed numerically and depend upon $\w$ and $\ell$. 
In~\cite{rigorous-results, methods-paper} the following exact expressions for the modes in~\eqref{Boulware-initial-data} were found:
\bea \psi^{\mathscr{I}^-} &=& e^{-i \w t_s} \left( \chi^\infty_L - \frac{E_L}{E_R} \chi^\infty_R \right)   \;, \nonumber \\
    \psi^{\mathscr{I}^+} &=& e^{-i \w t_s} \left( \chi^\infty_R - \frac{F_R}{F_L} \chi^\infty_L \right)   \;, \nonumber \\
    \psi^{\mathscr{H}^-} &=& e^{-i \w t_s} \left(\chi^\infty_L - \frac{E_L}{E_R} \chi^\infty_R \right)\;, \nonumber \\
    \psi^{\mathscr{H}^+} &=& e^{-i \w t_s} \frac{\chi^\infty_L}{F_L} \label{Boulware-modes} \;, \eea 

As shown in~\cite{methods-paper}, it is possible to expand the {\it in} modes in terms of $f^{\mathscr{I}^+}$ and $f^{\mathscr{H}^+}$ using matching coefficients obtained from the scalar products of the modes  $f^{\mathscr{I}^+}$ and  $f^{\mathscr{H}^+}$ with the {\it in} modes on a Cauchy surface containing the null shell trajectory $v = v_0$, the part of the future horizon with $v < v_H = v_0 - 4M$, and the part of past null infinity with $v > v_0$.  
The expansion takes the general form
\be
f^{in}_{\w \ell m} =  \frac{Y_{\ell m}}{r \sqrt{4 \pi}} \int_{0}^{\infty} \frac{d \w'}{\sqrt{\w'}}
\Big[ A^{\mathscr{I^+}}_{\w \w' \ell} \psi^{\mathscr{I}^+}_{\w'\ell} + B^{\mathscr{I^+}}_{\w \w' \ell} (\psi^{\mathscr{I}^+}_{\w' \ell})^{*}
 + A^{H^+}_{\w \w' \ell} \psi^{H^+}_{\w' \ell} + B^{H^+}_{\w \w'\ell} (\psi^{H^+}_{\w' \ell})^{*}\Big]\;. \label{General-in-modes}
\ee
General formulas for the matching coefficients are given in~\cite{methods-paper}.  For $\ell = 0$, a simplification can be made.  In the flat-space region inside the null shell the modes are
\be f^{\rm in}_{\w 0 0} = (f^{\rm in}_{\w 0 0})_v + (f^{\rm in}_{\w 0 0})_u = \frac{Y_{00}}{r \sqrt{4 \pi \w}} e^{-i \w v} - \frac{Y_{00}}{r \sqrt{4 \pi \w}}  e^{-i \w u}  \;. \label{fin-flat} \ee 
Because the expansion of $f^{\rm in}$ in~\eqref{General-in-modes} is linear in the expansion coefficients, one can find separate expansions of this form for $(f^{\rm in}_{\w 00})_v$ and $(f^{\rm in}_{\w 0 0})_u$.  The matching coefficients for these expansions are given in~\cite{quasinormal}.   

The Unruh state consists of the union of the $f^{\mathscr{I}^-}$ modes and Kruskal modes, $f^K$, that have the following initial data on the past horizon:
\be \psi^K_{\w \ell} = e^{-i \w U(u_s)}  \;, \label{psiK-def} \ee
with 
\be U = - 4 M e^{-\frac{u_s}{4 M}}  \;, \label{U-def} \ee
the outgoing radial Kruskal coordinate.
These modes can be expanded in terms of the $f^{\mathscr{H}^-}$ modes.  The expansion coefficients are obtained by computing the usual scalar product between the $f^K$ modes and the $f^{\mathscr{H}^-}$ modes on the past horizon.  Because the effective potential in the mode equation vanishes on this surface, the expansion coefficients are independent of the values of $\ell$ and $m$.   The result is
\be f^K_{\w \ell m} = \int_0^\infty d \w^\prime [\alpha^K_{\w \w'} f^{\mathscr{H}^-}_{\w' \ell m} + \beta^K_{\w \w'} f^{\mathscr{H}^- \; *}_{\w' \ell m}] \;. \label{fK-expansion} \ee
with~\cite{quasinormal}
\bes \bea \alpha^K_{\w_K \w'} &=&  \frac{1}{2 \pi} \sqrt{\frac{\w'}{\w_K}} \, (4M)^{1 +i 4 M \w'} \frac{\Gamma(\delta - i 4 M \w')}{ (-i \w_K + \epsilon)^{-i 4 M \w'}} \;, \label{alphaK} \\
\beta^K_{\w_K \w'} &=&  \frac{1}{2 \pi} \sqrt{\frac{\w'}{\w_K}} \, (4M)^{1 -i 4 M \w'} \frac{\Gamma(\delta + i 4 M \w')}{ (-i \w_K + \epsilon)^{i 4 M \w'}} \;. \label{betaK}
\eea \label{alphaK-betaK} \ees
Here $0<\delta \ll 1$ and $0<\epsilon \ll 1$ are integrating factors.

An alternative state~\cite{Unruh-1976}, which gives the same answer for the vacuum expectation value of quantities such as $\la \phi^2 \ra$ and $\la T_{ab} \ra$, results from the replacement of the $f^K$ modes with modes that we call $(f^{\mathscr{H}^-})_{\rm th}$.  These are proportional to $f^{\mathscr{H}^-}$ and can be written in the form~\cite{SdS-CQG}
\be (f^{\mathscr{H}^-}_{\w \ell m})_{\rm th} = \frac{e^{2  \pi M \w}}{[2 \sinh (4  \pi M |\w|)]^{1/2}} f^{\mathscr{H}^-}_{\w \ell m} \;.  \label{fth-def} \ee

The expansions of the quantum field $\phi$ in Schwarzschild spacetime for the {\it in} state, the Unruh state, and its alternate state are
\bes \bea \phi(x) &=& \sum_{\ell \, m} \int_0^\infty d \w [a^{\rm in}_{\w \ell m} f^{\rm in}_{\w \ell m} + a^{{\rm in} \; \dagger}_{\w \ell m} f^{{\rm in} \; *}_{\w \ell m} ] \;, \label{in-state} \\
&=& \sum_{\ell \, m} \int_0^\infty d \w [a^{\mathscr{H}^-}_{\w \ell m} f^{\mathscr{H}^-}_{\w \ell m} + a^{\mathscr{H}^- \; \dagger}_{\w \ell m} f^{\mathscr{H}^- \; *}_{\w \ell m} ] \nonumber \\
 & & + \sum_{\ell \, m} \int_0^\infty d \w_K [a^K_{\w_K \ell m} f^K_{\w_K \ell m} + a^{K \; \dagger}_{\w_K \ell m}] f^{K \; *}_{\w_K \ell m} \;, \label{Unruh-state} \\
 &=& \sum_{\ell \, m} \int_0^\infty d \w [a^{\mathscr{H}^-}_{\w \ell m} f^{\mathscr{H}^-}_{\w \ell m} + a^{\mathscr{H}^- \; \dagger}_{\w \ell m} f^{\mathscr{H}^- \; *}_{\w \ell m} ] \nonumber \\
 && + \sum_{\ell \, m} \int_{-\infty}^\infty d \w [a^{\rm th}_{\w \ell m} f^{\rm th}_{\w \ell m} + a^{{\rm th} \; \dagger}_{\w \ell m} f^{{\rm th} \; *}_{\w \ell m}] \;. \label{Unruh-alt-state}
\eea \ees 

\subsection{Method to solve the mode equation directly for the  in state modes}

The collapse modes carry waves incident from past null infinity and reflected at the regular center. Their data on the shell follow from Eq.~\eqref{fin-flat}. We compute their propagation through the Schwarzschild potential by solving a characteristic (Goursat) problem. The approach we take is to numerically solve the scattering problem from past null infinity to future null infinity, similar to the recent numerical studies in Refs.~\cite{FrauendienerStevensThwala2025, BarrerEtAl2026, DuarteBaptista2026, Zenginoglu2026Scattering, DemirbogaZenginoglu2026, HeTianZhang2026}. 

At each frequency, we subtract the stationary incoming mode and define the collapse correction
\begin{equation}
 D_\w=\psi^{\rm in}_{\w 0}-\psi^{\mathscr I^-}_{\w 0}.
\end{equation}
The flat-space outgoing coordinate $u$ labels the outgoing rays. With $J=du/du_s=(v_H-u)/(v_0-u)$, Eq.~\eqref{psi-eq} gives
\begin{equation}
 \partial_u\partial_vD_\w+\frac{V_{\rm eff}(r)}{4J}D_\w=0.
\end{equation}
Regularity at the center fixes the shell data:
\begin{equation}
 D_\w(u,v_0)=e^{-i\w v_0}-e^{-i\w u}
 -\psi^{\mathscr I^-}_{\w 0}(u_s(u),v_0).
\end{equation}
The common incoming component cancels in $D_\w$, leaving no independent incoming component at $\mathscr I^-$. The reference mode follows from the radial equation~\eqref{chi-eq} at the same frequency.

We integrate the characteristic equation in advanced time using high-order spatial quadrature and adaptive time integration. The data at a finite past boundary include the asymptotic scattering contribution from the omitted domain. We check convergence by refining the grid and extending the past domain. Adding the reference mode recovers $\psi^{\rm in}_{\w 0}$ at finite radius.

For subsequent evolution, we use stationary hyperboloidal scri-fixing, introduced in Ref.~\cite{Zenginoglu2008}, combining a suitable time transformation, spatial compactification, and conformal rescaling. Early applications evolved scalar fields~\cite{Zenginoglu2008Tails} and gravitational perturbations~\cite{ZenginogluNunezHusa2009,Zenginoglu2010Asymptotics} on Schwarzschild spacetime. Within this class, we choose a variant of the minimal gauge~\cite{AnsorgMacedo2016,MacedoJaramilloAnsorg2018,Macedo2020,Macedo2024}:
\begin{equation}
 r=\frac{2M}{1-\rho},\qquad t_s=2M[\tau-h(\rho)],\qquad
 h(\rho)=\log\rho+\log(1-\rho)-\frac{1}{1-\rho}.
\end{equation}
These slices cross the future horizon at $\rho=0$ and reach future null infinity at $\rho=1$, forming a bridge between the two null boundaries~\cite{Zenginoglu2025}. With $v_0=4M$, the slice $\tau_0=0.5$ lies entirely after the shell. The characteristic equations supply both the field and its time derivative there. Subtracting the common incoming phase gives $D_\w$ a finite outgoing limit at null infinity. We evaluate endpoint values and derivatives from their asymptotic expansions.

\subsection{Vacuum Polarization and Luminosity}

The symmetric two-point function is 
\be G^{(1)}(x,x') = \la \{\phi(x), \phi(x') \} \ra \;. \ee
If point splitting is used, the renormalized vacuum polarization is
\be \la \phi^2(x) \ra_{\rm ren} = \frac{1}{2} \lim_{x' \to x} [ G^{(1)}(x,x') - G^{(1)}_{\rm DS}(x,x')]  \;, \ee
where the second term is obtained from the DeWitt-Schwinger expansion~\cite{Christensen-1976}.   For a given quantum field, in the limit that the points come together,  $G^{(1)}_{\rm DS}(x,x')$ only depends on the spacetime geometry near the points.  It is thus independent of both the state and the global properties of the spacetime.  It is for this reason that outside the null shell trajectory 
\be \Delta \la \phi^2(x) \ra  = \la \phi^2(x) \ra_{\rm in} -  \la \phi^2(x) \ra_U  
\equiv \frac{1}{2} \lim_{x' \to x} [ G^{(1)}_{\rm in}(x,x') - G^{(1)}_{U}(x,x')] \label{Delta-phi2-def} \ee
is finite. 

The luminosity is given by the quantity $L = 4 \pi r^2 \la T_{u_s \, u_s} \ra $ evaluated at future null infinity. The difference in the luminosity between the {\it in} and Unruh states is
\be \Delta L = L_{\rm in}-L_U =  \lim_{\substack{t_s \to \infty \\ r\to \infty \\ u_s\; {\rm fixed}}} 4 \pi r^2 \lim_{x'\to x} \frac{1}{2}\partial_{u_s}\partial_{u'_s}\Big\{G_{in}^{(1)}(x,x')- G_U^{(1)}(x,x')\Big\}        \; . \ee

\section{Calculation of $\Delta \la \phi^2 \ra$ and $\Delta L$}

In what follows, we focus on two methods that we used to calculate the contribution of the spherically symmetric modes to $\Delta \la \phi^2 \ra$.  The late-time method was also used to compute the contribution of these modes to $\Delta L$.

\subsection{Early time calculation}

For a fixed and not too large radius $r$, the calculation of $\Delta \la \phi^2 \ra$  at relatively early times $t_s$ after the shell has crossed the radius $r$ can be broken into three parts: $0 < \lambda \ll 1 \ll \Lambda$. The first part is 
\be \Delta \la \phi^2 \ra_a = \int_0^\lambda d \w \Big\{ |f^{\rm in}_{\w}|^2 - |f^{\mathscr{I}^-}_\w|^2 - |f^{\rm th}_\w|^2\Big\} \;. \label{Dphi2-early-a} \ee
For this part, the integrand is computed by fitting modes with $\w \approx \lambda$ which have been computed numerically.  This is done for each value of $t_s$ and $r$.  The result is a relatively small contribution to $\Delta \la \phi^2 \ra$. It is important to note that each term above is infrared finite.  For $\ell = 0$, the $f^{\rm in}$ modes can be split into two parts as is discussed in Sec.~\ref{sec:chi-solns}.  However, the two parts are separately infrared divergent so it is important for small values of $\w'$ to combine them before computing the integral.

The second part is 
\be \Delta \la \phi^2 \ra_b = \int_\lambda^\Lambda d \w \Big\{|f^{\rm in}_{\w}|^2 - |f^{\mathscr{I}^-}_\w|^2 
- |f^{\rm th}_\w|^2 \Big\} \;.  \label{Dphi2-early-b} \ee
Here $\Lambda$ is large enough that each type of mode is approximately equal to the equivalent mode when the effective potential is set to zero.  The modes are all computed numerically.

The third part is 
\bea \Delta \la \phi^2 \ra_c &=& \frac{1}{2}\lim_{t_s^\prime \to t_s} \int_\Lambda^\infty d \w  \Big\{f^{\rm in}_{\w}(t_s,r) 
f^{{\rm in}\, *}_\w (t_s^\prime,r) - f^{{\mathscr{I}^-}}_\w (t_s,r) f^{\mathscr{I}^-\; *}_\w (t_s^\prime,r)  \nonumber \\  
& & \qquad \qquad \qquad  - f^{\rm th}_\w (t_s,r)  f^{{\rm th} \, *}_\w (ts^\prime,r)  + {\rm cc} \Big\} \;,  \label{Dphi2-early-c-1}\eea
where cc denotes the complex conjugate.  If  $\Lambda$ is large enough, then large frequency approximations for the mode functions can be used and the thermal factor $\coth\left( \frac{\pi \w}{\kappa} \right)$ can be expanded in powers of $e^{- \pi \w/\kappa}$. To lowest order, the thermal factor is equal to $1$ and the large frequency approximations are just the no-scattering solutions that are obtained with the same initial data but no effective potential in the mode equation.  In that case
\bea \Delta \la \phi^2 \ra_c &=&  \lim_{t_s^\prime \to t_s} \frac{|Y_{00}|^2}{4 \pi r^2} \int_\Lambda^\infty \frac{d \w}{\w} \left\{  -\cos[\w (v - u(u_s^\prime)] - \cos[\w(v^\prime -u(u_s)] + \cos[\w (u(u_s) - u(u_s^\prime)] \right. \nonumber \\
& & \left. - \cos[\w (u_s - u_s^\prime)] \right\} \nonumber \\
  &=&  \frac{1}{16 \pi^2 r^2} \left\{  2 {\rm ci}[\Lambda(v-u(u_s)] - \lim_{t_s^\prime \to t_s} \left[ {\rm ci}\{\Lambda |u(u_s)-u(u_s^\prime)| \} - {\rm ci}\{\Lambda (t_s - t^\prime_s) \} \right] \right\} \nonumber \\
  &=& \frac{1}{16 \pi^2 r^2} \left\{  2 {\rm ci}[\Lambda(v-u(u_s)] - \log \frac{du}{d u_s} \right\} = 
   \frac{1}{16 \pi^2 r^2} \left\{  \log\left|\frac{v_0-u}{v_H - u} \right| + 2 {\rm ci}[\Lambda(v-u(u_s)] \right\} \;.
\nonumber \\   \label{Dphi2-early-c-2}
\eea

\subsection{Late-time calculation}

The calculation of $\Delta \la \phi^2 \ra$ for fixed $r$ at late times is done using a different method.  One reason is that at late enough times for fixed $\w$ and $r$, $f_u \to - f^K$. A second reason is that at late times one must go to much larger values of $\w$ before the no-scattering solutions for $f_u$ and $f^K$ become good approximations.  The values increase exponentially in $t_s$.  For the Kruskal modes there is a scaling that can be used to transform modes at early times to modes at late times if $r$ is fixed.  The relationship, can be derived by examining the Bogolubov coefficients $\alpha_{\w \w'}$ and $\beta_{\w \w'}$ and the expansion of $f^K$ in terms of them.  One finds from~\eqref{fK-expansion} and~\eqref{alphaK-betaK} that for fixed $r, \theta, \phi$,
 the quantity $\sqrt{\w} f^K_{\w \ell m}(t_s,r, \theta,\phi)$ is the same for values of $t_s$ and $\w$ that give the same value of 
 \be z = t_s - 4 M \log [ 4 M \w] \; . \label{z-def} \ee

Specifically, we replace~\eqref{Dphi2-early-a} by
\be \Delta \la \phi^2 \ra_a = \int_0^\lambda d \w [ |f^{\rm in}_{\w}|^2 - |f^{\mathscr{I}^-}_\w|^2 - |f^{K}_\w|^2] \;.
,\label{Dphi2-early} \ee
and~\eqref{Dphi2-early-b} by 
\be \Delta \la \phi^2 \ra_b = \int_\lambda^\Lambda d \w [ |(f^{\rm in}_{\w})_v|^2 - |f^{\mathscr{I}^-}_\w|^2 + |(f^{\rm in}_{\w})_u|^2  - |f^K_\w|^2 + 2 \Re\{(f^{\rm in}_{\w})_v f^{K \, *}_\w \} ] \;. \label{Dphi2-late} \ee
Eq.~\eqref{Dphi2-early-c-1} becomes
\bea \Delta \la \phi^2 \ra_c &=& \lim_{t_s^\prime \to t_s} \int_\Lambda^\infty d \w \frac{1}{2} [(f^{\rm in}_{\w})_v (t_s,r) 
(f^{{\rm in}\, *}_\w)_v (t_s^\prime,r) - f^{{\mathscr{I}^-}}_\w (t_s,r) f^{\mathscr{I}^-\; *}_\w (t_s^\prime,r)   
\nonumber \\
& & + (f^{\rm in}_{\w})_u (t_s,r) 
(f^{{\rm in}\, *}_\w)_u (t_s^\prime,r)  - f^K_\w (t_s,r)  f^{K \, *}_\w (ts^\prime,r) +  {\rm cc} ] \nonumber \\
 && +  2 \int_\Lambda^\infty d \w   \Re\{(f^{\rm in}_{\w})_v (f^{\rm in \, *}_\w)_u  \}  + {\rm cc}     \;.  \label{Dphi2-late-c-1} \eea
Note that the last term does not have the points split because it is finite.  If $\Lambda$ is large enough that the no-scattering modes can be used then one finds
\bea \Delta \la \phi^2 \ra_c &=&  \lim_{t_s^\prime \to t_s} \frac{|Y_{00}|^2}{4 \pi r^2} \int_\Lambda^\infty \frac{d \w}{\w}  \left\{  \cos[\w (u(u_s) - u(u_s^\prime))] -  \cos[\w (U(u_s) - U(u_s^\prime))]   \right. \nonumber \\
& & \left.  - 2\cos[\w (v - u(u_s)]  \right\} \nonumber \\
 &=&  \frac{1}{16 \pi^2 r^2} \left\{  - \lim_{t_s^\prime \to t_s} \left[ {\rm ci}\{\Lambda |u(u_s)-u(u_s^\prime)| \} - {\rm ci}\{\Lambda (U(t_s) - U(t^\prime_s)) \} \right] +  2 {\rm ci}[\Lambda(v-u(u_s)] \right\} \nonumber \\
 &=& \frac{1}{16 \pi^2 r^2} \left\{  - \log \frac{du}{d u_s} + \log \frac{dU}{d u_s} + 2 {\rm ci}[\Lambda(v-u(u_s)] \right\} \nonumber \\ 
 &=& 
   \frac{1}{16 \pi^2 r^2} \left\{ \log\left|\frac{v_0-u}{v_H - u}\right| -   \frac{u_s}{4 M}  + 2 {\rm ci}[\Lambda(v-u(u_s)] \right\} \;.
   \label{Dphi2-late-c-2}
\eea

The calculation of the corresponding contribution to the luminosity proceeds in a similar way.  The primary difference is that since the calculation is done at $\mathscr{I}^+$, we can evaluate the expressions for the Boulware modes in~\eqref{Boulware-modes} by setting $\psi^\infty_L$ to zero\footnote{Technically on $\mathscr{I}^+$, $\psi^\infty_L \to \lim_{v \to \infty} e^{-i \w v}$.  This is effectively zero in the sense that wave packets with well-behaved coefficients and a finite range of values of $\w$ will vanish in this limit. } and $\psi^\infty_R = e^{-i \w u_s}$. The result is 

\bea \Delta L_c &=&  -\lim_{u_s^\prime \to u_s} |Y_{00}|^2 \int_\Lambda^\infty d \w  \;\w \Big\{ \frac{du}{du_s}\frac{du'}{du'_s} \cos[\w (u(u_s) - u(u_s^\prime))] -  \frac{dU}{du_s}\frac{dU'_s}{du_s'}\cos[\w (U(u_s) - U(u_s^\prime))]   \Big\} \nonumber \\ 
 &=& \frac{1}{4\pi} \Big\{\lim_{u_s^\prime \to u_s} \frac{(v_H-u(u_s))(v_H-u(u_s'))}{(v_0-u(u_s))(v_0-u(u_s'))}\Bigg[\frac{\cos{\big(\Lambda(u(u_s)-u(u_s'))\big)}}{\big(u(u_s)-u(u_s')\big)^2}
 +\frac{\sin{\big(\Lambda(u(u_s)-u(u_s'))\big)}}{u(u_s)-u(u_s')}\Bigg]\nonumber \\ &&
 -e^{-\frac{(u_s+u_s')}{4M}}\Bigg[\frac{\cos{\big(\Lambda(U(u_s)-U(u_s'))\big)}}{\big(U(u_s)-U(u_s')\big)^2}
 +\frac{\sin{\big(\Lambda(U(u_s)-U(u_s'))\big)}}{U(u_s)-U(u_s')}\Bigg]\Big\}\nonumber \\
 &=&\frac{1}{768\pi M^2}-\frac{M(v_H-u+M)}{3\pi(v_0-u)^4}+\frac{\Lambda^2}{8\pi}\Big\{(\frac{v_H-u}{v_0-u})^2-e^{\frac{u_s}{2M}}\Big\}.
   \label{delta-luminosity-large-w}
\eea

\section{Numerical Results}

\begin{figure*}[t]
    \centering
    \begin{minipage}[t]{0.50\textwidth}
        \centering
        \includegraphics[width=\linewidth]{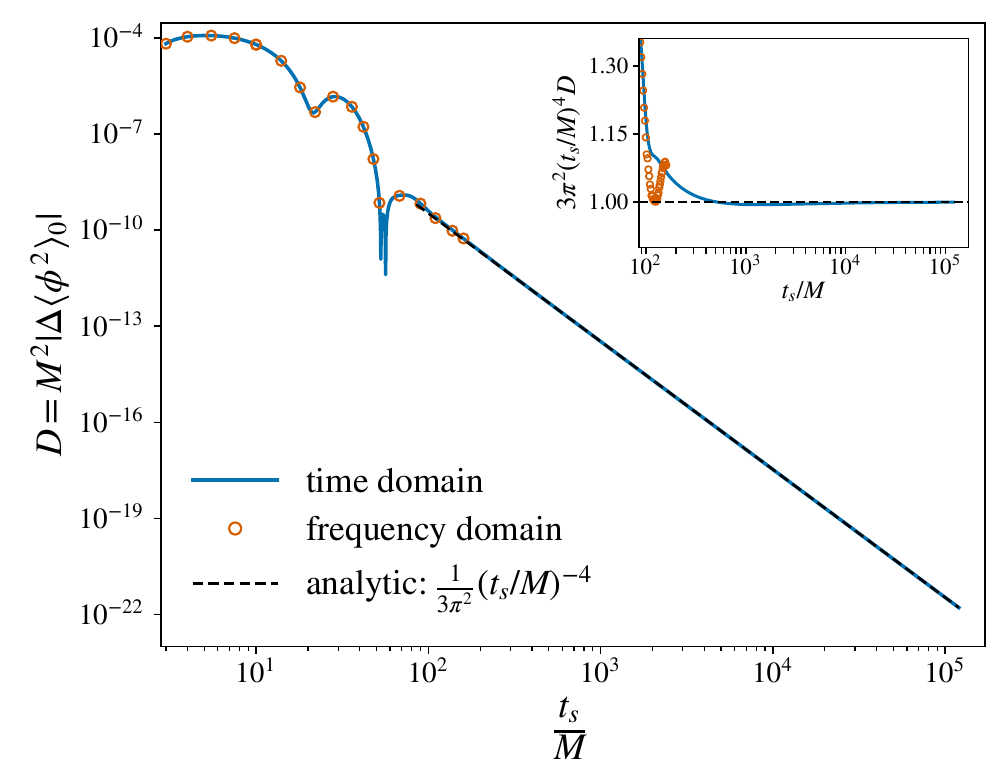}\\[-0.5ex]
        (a)
    \end{minipage}\hfill
    \begin{minipage}[t]{0.50\textwidth}
        \centering
        \includegraphics[width=\linewidth]{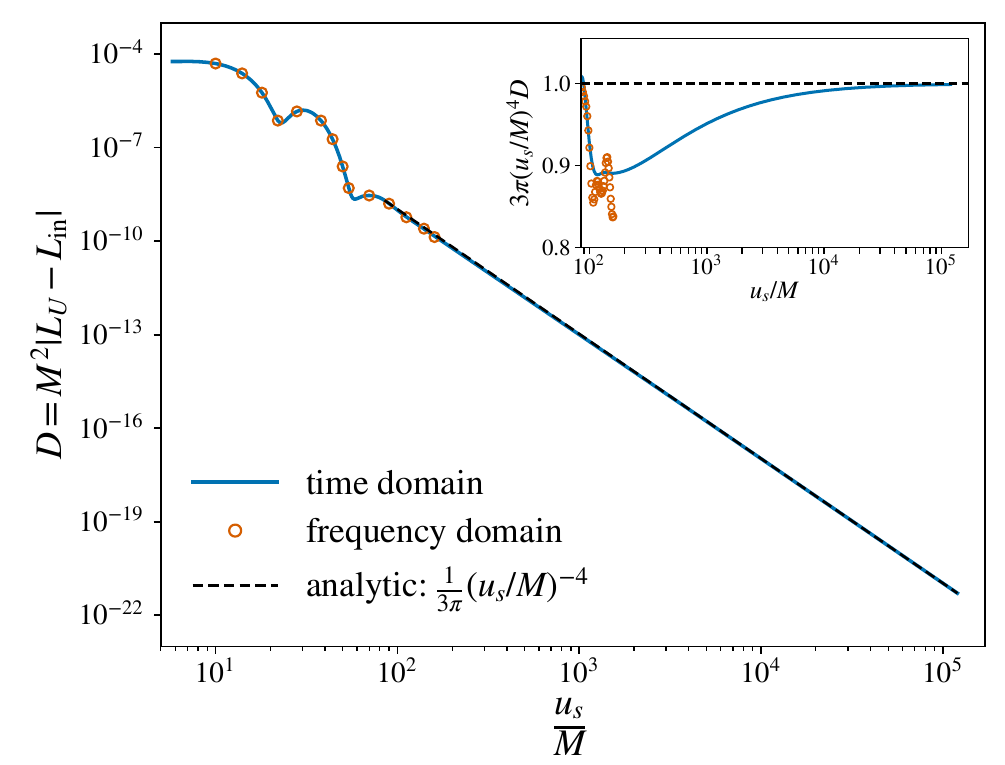}\\[-0.5ex]
        (b)
    \end{minipage}
    \caption{Relaxation of monopole observables after null-shell collapse: (a) the vacuum-polarization difference $D=M^2|\Delta\la\phi^2\ra_0|$ at $r=3M$; (b) the luminosity difference $D=M^2|L_U-L_{\mathrm{in}}|$ at future null infinity.
    Blue lines show the time-domain calculations with hyperboloidal evolution; orange open circles show the independent frequency-domain calculations.
    Dashed lines are the analytic predictions $(3\pi^2)^{-1}(t_s/M)^{-4}$ in (a) and $(3\pi)^{-1}(u_s/M)^{-4}$ in (b), with no fitted parameters.
    Insets show the ratios to these predictions, approaching unity at late times.}
    \label{fig:Phi2}
    \label{fig:L}
\end{figure*}

The quantity $\Delta \la \phi^2 \ra$ has been computed for $3\le t_s/M\le 1.2\times10^5$ for the case $r = 3M$.  The results are shown in Fig.~\ref{fig:Phi2}(a).  At intermediate times the difference is dominated by the quasinormal mode with the smallest complex frequency which is equal to  $0.1105 - 0.1049 i$~\cite{quasi-review}.   At late times the difference is dominated by a tail term of the form $t_s^{-p}$ with $p \approx 4$.

The quantity $\Delta L$ is shown in Fig.~\ref{fig:L}(b).  The outgoing radial null coordinate $u_s$ can be used to parameterize future null infinity.  It  is seen that $\Delta L$ is again dominated by the quasinormal mode with the smallest complex frequency for intermediate values of $u_s$.  For large values, there is a tail term which decays like $u_s^{-4}$.  The results of an analytic calculation of this tail in Sec.~\ref{sec:luminosity-tail} are also given.

\section{Origin of the late-time tails}
\label{sec:luminosity-tail}

The power-law transients in the luminosity and vacuum polarization come from curvature scattering of the distant collapse history. We calculate their leading coefficients after integrating over the incoming-state frequency. Write $\bar u=u_s/(2M)$ for the observation time and $w=u/(2M)$ for the affine shell coordinate. The outgoing affine vacuum data have reduced covariance $-\log|w-w'|$, up to an additive constant removed by the luminosity derivatives. With $v_H=0$, the shell map is
\begin{equation}
 y=w-2\log(-w/2),\qquad y=\frac{u_s(u)}{2M},\qquad w<0.
 \label{eq:tail-shell-map}
\end{equation}
The dimensionless luminosity difference, with the mode normalization used above, is
\begin{equation}
 F(\bar u)=16\pi M^2\Delta L_0.
 \label{eq:tail-flux-normalization}
\end{equation}

The low-frequency reflection phase fixes the late response. Define $\nu=2M\omega'$ and $\mathcal R_\nu=E_L/E_R=e^{2i\delta_0(\nu)}$, so the reflected wave has the minus sign in Eq.~\eqref{Boulware-modes}. For $\nu>0$, the monopole phase is~\cite{CorreiaEtAl2025}
\begin{equation}
 \delta_0(\nu)=\nu\left[\log(2|\nu|)+\gamma_{\rm E}-\frac12\right]
 +\left(\frac{11\pi}{12}+i\right)\nu^2+\cdots ,
 \label{eq:tail-reflection-phase}
\end{equation}
where $\gamma_{\rm E}$ is Euler's constant. Extend $\mathcal R_{-\nu}=\overline{\mathcal R_\nu}$ and define
\begin{equation}
 c(t)=\frac{1}{2\pi}\int_{-\infty}^{\infty}
       e^{-i\nu t}\mathcal R_\nu\,d\nu .
\end{equation}
The logarithmic frequency term gives $c(t)\sim-t^{-2}$ at positive delay.

Let $s(\bar u,w)$ be the shell-to-infinity response density, with the minus sign of the outgoing shell data in Eq.~\eqref{fin-flat} factored out. Its stationary counterpart is $c(\bar u-w)$. In the distant-source region $w=-\bar u a$, $a>0$, their leading terms coincide:
\begin{equation}
 s(\bar u,-\bar u a)\sim c\bigl(\bar u(1+a)\bigr)
 \sim-\frac{1}{\bar u^2(1+a)^2}.
\end{equation}
This equality cancels the candidate cubic luminosity term between the shell contribution and the stationary incoming past removed by the Unruh subtraction. Expanding the radial solution and the shell map through the next order gives
\begin{equation}
 s(\bar u,-\bar u a)-c\bigl(\bar u(1+a)\bigr)
 \sim\frac{f(a)}{\bar u^3}, \qquad 
 f(a)=\frac{2\log a-4\log(1+a)+4}{(1+a)^3}.
 \label{eq:tail-response-defect}
\end{equation}
The constant $4$ includes the second-order reflection phase in Eq.~\eqref{eq:tail-reflection-phase}. The identity $\int_0^\infty f(a)\,da=0$ preserves the cancellation of the response masses.

At fixed $w$, the differentiated response difference has leading term $\bar u^{-4}[-3f(a)-af'(a)]$. Differentiating the affine covariance against the local reflected response supplies the factor $1/(\bar u-w)$. After combining the collapse and reference sectors, the leading cross term gives the matched-asymptotic prediction
\begin{equation}
 F(\bar u)
 \sim\frac{2}{\bar u^4}\int_0^\infty
       \frac{3f(a)+af'(a)}{1+a}\,da
 =\frac{2}{\bar u^4}\int_0^\infty
       \frac{(2+3a)f(a)}{(1+a)^2}\,da
 =-\frac{1}{3\bar u^4}.
 \label{eq:tail-coefficient-integral}
\end{equation}
The second line follows by integration by parts; the convergent logarithmic moments evaluate to $-1/3$. Restoring physical units yields
\begin{equation}
 M^2\Delta L_0
 \sim-\frac{1}{3\pi}\left(\frac{u_s}{M}\right)^{-4}.
 \label{eq:luminosity-quartic-prediction}
\end{equation}
Thus the collapse luminosity approaches the Unruh value from below at late times. The coefficient follows from the shell geometry and the Schwarzschild scattering expansion, independently of a fit.

The same leading covariance gives the vacuum polarization at fixed radius. Put $\bar t=t_s/(2M)$ and $\bar r=r/(2M)$. Matching the horizon-regular zero-frequency radial solution $\bar r$ to unit incoming amplitude in Eq.~\eqref{Boulware-modes} gives
\begin{equation}
 \chi_L^\infty-\mathcal R_\nu\chi_R^\infty
 =-2i\nu\bar r+O(\nu^2\log|\nu|),\qquad r>2M\ \text{fixed}.
 \label{eq:tail-fixed-radius-transfer}
\end{equation}
The outgoing radiation amplitude tends to $-\mathcal R_0=-1$. Thus the leading distant-history contribution at fixed radius is obtained by applying $-2\bar r\partial_{\bar t}$ in each argument of the radiation covariance. These are the same independent derivatives entering $F$. For the reduced covariance $8\pi^2rr'\Delta G^{(1)}_0$, the radial factors give $4\bar r^2F(\bar t)$ at coincidence. Assuming the remaining matched terms are subleading at fixed radius, Eq.~\eqref{Delta-phi2-def} gives
\begin{equation}
 M^2\Delta\la\phi^2\ra_0(t_s,r)
 \sim\frac{4\bar r^2}{64\pi^2\bar r^2}F(\bar t)
 \sim-\frac{1}{48\pi^2\bar t^4} =-\frac{1}{3\pi^2}\left(\frac{t_s}{M}\right)^{-4}.
 \label{eq:phi2-quartic-prediction}
\end{equation}
The radial factors cancel, so the leading coefficient is independent of the chosen fixed radius. The bounded difference between $\bar t$ and $\bar u$ changes only subleading terms.

\section{Discussion}

We have explored the approach of the vacuum polarization and the luminosity to their Unruh state values for a massless minimally coupled scalar field in a spacetime where a null shell collapses to form a black hole.  For each quantity, this has been done by computing the contribution of the spherically symmetric modes to the difference between the value of the quantity for the {\it in} vacuum state and its value for the Unruh state. 

To do the computations we have used two completely different methods.  The frequency domain method involves expanding the modes for the {\it in} state and Unruh state in terms of complete sets of Boulware modes which can be computed using separation of variables.  The time domain method involves directly solving the mode equation~\eqref{psi-eq} which is a partial differential equation.  We find good agreement between the two methods.

The results show the late-time approach to the Unruh state of the vacuum polarization and the luminosity.  This approach gives quantitative information about how quickly information about how the black hole formed disappears from the region outside the event horizon when backreaction effects are not taken into account.  

At intermediate times, the vacuum polarization and luminosity show exponentially damped oscillations. Their late-time tails are consistent with $t_s^{-4}$ and $u_s^{-4}$, respectively.  Analytic calculations in Sec.~\ref{sec:luminosity-tail} identify the curvature scattering of the collapse history as the source of the late-time relaxation.  The cubic terms in the contributions from the collapse and the Unruh state cancel.  The fixed-frequency mode tails found in Ref.~\cite{quasinormal} enter the frequency integral together with interference terms.

These results show unambiguously that the approach to the Unruh state for the vacuum polarization and luminosity is significantly slower than it is for a null shell that collapses to form a black hole in two dimensions where the approach is exponential in $t_s$ and $u_s$ respectively.  It seems very likely that the qualitative features of a quasinormal mode period followed by a tail term will be universal for the approach to the Unruh state of quantities such as the stress-energy of the Hawking radiation when a black hole forms from collapse. It is also likely that the coefficients of the quasinormal modes and the tail term will carry information about how the black hole formed.  

\acknowledgments 

P. R. A. would like to thank Adam Levi for sharing some numerical data on solutions to the mode equation and helpful conversations about it, and Raymond Clark for providing power series solutions to the radial mode equation in Schwarzschild spacetime.  P. R. A. would also like to thank Northwestern University's Center for Interdisciplinary
Exploration and Research in Astrophysics (CIERA) for
hosting him during a visit in August of 2023. This work was supported in part by the National Science Foundation under Grants  No. PHY-1505875, PHY-1912584, and PHY-2309186 to Wake Forest University and Grant No. 2309084 to the University of Maryland.  A.F. acknowledges partial financial support by the Spanish Grant PID2023-149560NB-C21 funded by MCIN/AEI/10.13039/501100011033
and FEDER, European Union, and the Severo Ochoa Excellence Grant
CEX2023-001292-S. 
Some of the numerical work was done using the WFU DEAC Cluster; we thank the WFU Provost's Office and Information Systems Department for their generous support.

\end{document}